\documentclass[aps,showpacs,prd]{revtex4}%
\usepackage{amssymb}
\usepackage{amsfonts}
\usepackage{amsmath}
\usepackage{graphicx}
\usepackage{color}
\usepackage[dvips]{epsfig}
\usepackage[dvips]{graphicx}%
\providecommand{\U}[1]{\protect\rule{.1in}{.1in}}

\begin{document}
\title{Aspects of a planar nonbirefringent and CPT-even electrodynamics stemming from
the Standard Model Extension}
\author{Rodolfo Casana, Manoel M. Ferreira Jr, Roemir P. M. Moreira}
\affiliation{Departamento de F\'{\i}sica, Universidade Federal do Maranh\~{a}o (UFMA),
Campus Universit\'{a}rio do Bacanga, S\~{a}o Lu\'{\i}s - MA, 65085-580, Brazil.}

\begin{abstract}
We have studied a (1+2)-dimensional Lorentz-violating model which is obtained
from the dimensional reduction of the nonbirefringent sector of the CPT-even
electrodynamics of the standard model extension (SME). The planar theory
contains a gauge sector and a scalar sector which are linearly coupled by
means of a Lorentz-invariance violating (LIV) vector, $S^{\mu}$, while the
kinetic terms of both sectors are affected by the components of a
Lorentz-violating symmetric tensor, $\kappa^{\mu\nu}$. The energy-momentum
tensor reveals that both sectors present energy stability for sufficiently
small values of the Lorentz-violating parameters. The full dispersion relation
equations are exactly determined and analyzed for some special configurations
of the LIV backgrounds, showing that the planar model is entirely
nonbirefringent at any order in the LIV parameters. At first order, the gauge
and scalar sectors are described by the same dispersion relations. Finally,
the equations of motion have been solved in the stationary regime and at first
order in the LIV parameters. It is observed that the Lorentz-violating
parameters do not alter the asymptotical behavior of the electric and magnetic
fields but induce an angular dependence which is not present in Maxwell's
planar theory.

\end{abstract}

\maketitle

\section{Introduction}

Since the establishment of the special theory of relativity as a true of
nature, Lorentz symmetry has been taken as a key ingredient of theoretical
physics. A motivation for studies involving the violation of Lorentz symmetry
is the demonstration that string theories may support spontaneous violation of
this symmetry \cite{Samuel}, with important and interesting connections with
the physics in the Planck energy scale. The Standard Model Extension (SME)
\cite{Colladay} has arisen as theoretical framework for addressing Lorentz
violation (LV) in a broader context than the usual Standard Model, in an
attempt of scrutinizing remanent effects of this violation in several low
energy systems. In this way, the SME incorporates Lorentz-violating
coefficients to all sectors of the standard model and to general relativity,
representing a suitable tool for investigating Lorentz violation in several
distinct respects.

The violation of Lorentz symmetry in the gauge sector of the SME is governed
by a CPT-odd and a CPT-even tensor, yielding some unconventional phenomena
such as vacuum birefringence and Cherenkov radiation. The LV coefficients are
usually classified in accordance with the parity and birefringence. The
CPT-odd term is represented by the Carroll-Field-Jackiw (CFJ) background
\cite{Jackiw}, which is also parity-odd and birefringent.\ This
electrodynamics has been much investigated, encompassing aspects as diverse
as: consistency aspects and modifications induced in QED
\cite{Adam,Soldati,Higgs}, supersymmetry \cite{Susy}, vacuum Cherenkov
radiation emission \cite{Cerenkov1}, finite-temperature contributions and
Planck distribution \cite{winder1, Tfinite}, electromagnetic propagation in
waveguides \cite{winder2}, Casimir effect \cite{Casimir}.

The CPT-even gauge sector of the SME is represented by the CPT-even tensor,
$\left(  k_{F}\right)  _{\alpha\beta\mu\nu},$ composed of 19 independent
coefficients, with nine nonbirefringent and ten birefringent ones. This sector
has been studied since 2002 \cite{KM1,KM2}, \cite{KM3}, \cite{Kostelec}, being
represented by the following Lagrangian:
\begin{equation}
\mathit{{\mathcal{L}}}_{(1+3)}=-\frac{1}{4}F_{\hat{\mu}\hat{\nu}}F^{\hat{\mu
}\hat{\nu}}-\frac{1}{4}\left(  k_{F}\right)  _{\hat{\mu}\hat{\nu}\hat{\lambda
}\hat{\kappa}}F^{\hat{\mu}\hat{\nu}}F^{\lambda\kappa}-J_{\hat{\mu}}A^{\hat
{\mu}},\label{L1}%
\end{equation}
where the indices with hat, $\hat{\mu},\hat{\nu},$ run from 0 to 3,
$A^{\hat{\mu}}$ is the four-potential and $F_{\hat{\mu}\hat{\nu}}$ is the
usual electromagnetic field tensor. The tensor $\left(  k_{F}\right)
_{\hat{\mu}\hat{\nu}\hat{\lambda}\hat{\kappa}}$ stands for the
Lorentz-violating coupling and possesses the symmetries of the Riemann tensor,
$\left(  k_{F}\right)  _{\hat{\mu}\hat{\nu}\hat{\lambda}\hat{\kappa}}=-\left(
k_{F}\right)  _{\hat{\nu}\hat{\mu}\hat{\lambda}\hat{\kappa}},$ $\left(
k_{F}\right)  _{\hat{\mu}\hat{\nu}\hat{\lambda}\hat{\kappa}}=-\left(
k_{F}\right)  _{\hat{\mu}\hat{\nu}\hat{\kappa}\hat{\lambda}},\ $\ $\left(
k_{F}\right)  _{\hat{\mu}\hat{\nu}\hat{\lambda}\hat{\kappa}}=\left(
k_{F}\right)  _{\hat{\lambda}\hat{\kappa}\hat{\mu}\hat{\nu}},$ $\left(
k_{F}\right)  _{\hat{\mu}\hat{\nu}\hat{\lambda}\hat{\kappa}}+\left(
k_{F}\right)  _{\hat{\mu}\hat{\lambda}\hat{\kappa}\hat{\nu}}+\left(
k_{F}\right)  _{\hat{\mu}\hat{\kappa}\hat{\nu}\hat{\lambda}}=0,$ and a double
null trace, $\left(  k_{F}\right)  ^{\hat{\mu}\hat{\nu}}{}_{\hat{\mu}\hat{\nu
}}=0.$ A very useful parametrization for addressing this electrodynamics is
the one presented in Refs. \cite{KM1,KM2}, in which the nineteen LV components
are enclosed in four $3\times3$ matrices, defined as%
\begin{align}
\left(  \kappa_{DE}\right)  ^{j\kappa}  & =-2\left(  k_{F}\right)
^{0j0\kappa},\text{ }\left(  \kappa_{HB}\right)  ^{j\kappa}\;=\;\frac{1}%
{2}\epsilon^{jpq}\epsilon^{\kappa lm}\left(  k_{F}\right)  ^{pqlm}%
,\label{P1}\\[0.3cm]
\text{ }\left(  \kappa_{DB}\right)  ^{j\kappa}  & =-\left(  \kappa
_{HE}\right)  ^{\kappa j}\;=\;\epsilon^{\kappa pq}\left(  k_{F}\right)
^{0jpq}.\label{P2}%
\end{align}
The matrices $\kappa_{DE},\kappa_{HB}$ contain together 11 independent
components while $\kappa_{DB},\kappa_{HE}$ possess together 8 components,
which sums the 19 independent elements of the tensor $\left(  k_{F}\right)
_{\alpha\nu\rho\varphi}$. \ The ten birefringent components are severely
constrained by astrophysical tests involving high-quality cosmological
spectropolarimetry data, which have yielded stringent upper bounds at the
level of 1 part in $10^{32}$ \cite{KM1,KM2} and 1 part in $10^{37}$\cite{KM3}.
The nonbirefringent components are embraced by the matrices $\widetilde{\kappa
}_{e-}$ (six elements) and $\widetilde{\kappa}_{o+}$ (three elements), and can
be constrained by means of laboratory tests \cite{Cherenkov2} and the absence
of emission of Cherenkov radiation by UHECR (ultrahigh energy cosmic rays)
\cite{Klink2,Klink3}. These coefficients also undergo restriction at the order
of 1 part in $10^{17}$ considering their sub-leading birefringent role
\cite{Exirifard}. This CPT-even sector has\ also been recently investigated in
connection with consistency aspects in Refs. \cite{Klink4,Fred}.

Planar theories have been investigated since the beginning of 80's
\cite{Deser}, and have gained much attention due its connection with the
Chern-Simons theories \cite{Dunne}, planar superconductivity, anyons and
quantum Hall effect \cite{Ezawa}, and planar vortex configurations
\cite{Vortex}. The great importance of these topics has amounted to a great
development for the planar theories \cite{Avinash}.

A\ CPT-even field theory in (1+2)-dimensions model with Lorentz violation was
recently attained by means of the dimensional reduction of the CPT-even gauge
sector of the Standard Model Extension \cite{DreducCPT}. The resulting planar
electrodynamics is composed of a gauge and scalar sectors, both endowed with
Lorentz violation, whose planar Lagrangian is
\begin{equation}
\mathit{{\mathcal{L}}}_{(1+2)}=-\frac{1}{4}F_{\mu\nu}F^{\mu\nu}-\frac{1}%
{4}Z_{\mu\nu\lambda\kappa}F^{\mu\nu}F^{\lambda\kappa}+\frac{1}{2}\partial
_{\mu}\phi\partial^{\mu}\phi-C_{\mu\lambda}{\partial^{\mu}\phi\partial
^{\lambda}\phi}+T_{\mu\lambda\kappa}{\partial^{\mu}\phi}F^{\lambda\kappa
},\label{LP1}%
\end{equation}
where $Z_{\mu\nu\lambda\kappa},C_{\mu\lambda},T_{\mu\lambda\kappa}$ are LIV
tensors which have together 19 components and present the following
symmetries:%
\begin{equation}
Z_{\mu\nu\lambda\kappa}=\;Z_{\lambda\kappa\mu\nu},\text{ }Z_{\mu\nu
\lambda\kappa}=-Z_{\nu\mu\lambda\kappa},\text{ \ }Z_{\mu\nu\lambda\kappa
}=-Z_{\mu\nu\kappa\lambda},
\end{equation}%
\begin{align}
Z_{\mu\nu\lambda\kappa}+Z_{\mu\lambda\kappa\nu}+Z_{\mu\kappa\nu\lambda}  &
=0,\label{Perm}\\[0.3cm]
T_{\mu\lambda\kappa}+T_{\lambda\kappa\mu}+T_{\kappa\mu\lambda}  & =0,\text{ }%
\end{align}%
\begin{equation}
C_{\mu\lambda}=C_{\lambda\mu},\text{ }T_{\mu\lambda\kappa}=-T_{\mu
\kappa\lambda}.\label{sym1}%
\end{equation}
Some aspects of this model, involving wave equations and dispersion relations,
were addressed in Ref. \cite{DreducCPT}, having shown that the pure abelian
gauge or electromagnetic sector presents nonbirefringence at any order. The
birefringence in this model is associated with the elements of the coupling
tensor, $T_{\mu\lambda\kappa}.$

In the present work, we accomplish the dimensional reduction of the
nonbirefringent gauge sector of the SME, represented by 9 components which can
be incorporated in a symmetric and traceless tensor $\kappa_{\hat{\nu}%
\hat{\rho}},$ defined as the contraction \cite{Altschul}: $\kappa_{\hat{\nu
}\hat{\rho}}=\left(  k_{F}\right)  _{\text{ \ }\hat{\nu}\hat{\alpha}\hat{\rho
}}^{\hat{\alpha}}.$ The\ nonbirefringent components of the tensor $\left(
k_{F}\right)  _{\hat{\mu}\hat{\nu}\hat{\lambda}\hat{\kappa}}$ are parametrized
as
\begin{equation}
\left(  k_{F}\right)  _{\hat{\mu}\hat{\nu}\hat{\lambda}\hat{\rho}}=\frac{1}%
{2}\left(  g_{\hat{\mu}\hat{\lambda}}\kappa_{\hat{\nu}\hat{\rho}}-g_{\hat{\mu
}\hat{\rho}}\kappa_{\hat{\nu}\hat{\lambda}}-g_{\hat{\nu}\hat{\lambda}}%
\kappa_{\hat{\mu}\hat{\rho}}+g_{\hat{\nu}\hat{\rho}}\kappa_{\hat{\mu}%
\hat{\lambda}}\right)  ,
\end{equation}
which implies
\begin{equation}
\left(  k_{F}\right)  _{\hat{\mu}\hat{\nu}\hat{\lambda}\hat{\rho}}F^{\hat{\mu
}\hat{\nu}}F^{\hat{\lambda}\hat{\rho}}=2\kappa_{\hat{\nu}\hat{\rho}}%
F_{\hat{\lambda}}^{\text{ \ \ }\hat{\nu}}F^{\hat{\lambda}\hat{\rho}},
\end{equation}
so that the Lagrangian (\ref{L1}) takes on the form
\begin{equation}
\mathit{{\mathcal{L}}}_{(1+3)}=-\frac{1}{4}F_{\hat{\mu}\hat{\nu}}F^{\hat{\mu
}\hat{\nu}}-\frac{1}{2}\kappa_{\hat{\nu}\hat{\rho}}F_{\hat{\lambda}}{}%
^{\hat{\nu}}F^{\hat{\lambda}\hat{\rho}}-J_{\hat{\mu}}A^{\hat{\mu}}.\label{L2}%
\end{equation}
Some properties of this nonbirefringent electrodynamics were investigated in
Ref.\cite{Fred}, in which the corresponding Feynman gauge propagator was
evaluated and some of its consistency properties (causality and unitarity)
were analyzed.

In the present work, we perform the dimensional reduction of Lagrangian
(\ref{L2}), which produces a nonbirefringent planar theory composed of 9 LIV
parameters instead of the 19 ones attained in Ref. \cite{DreducCPT}. In this
simpler framework, Lorentz violation is controlled only by a rank-2, which
modifies the kinetic part of the scalar and gauge sectors, and a rank-1
tensor, which couples both sectors. The density of energy was evaluated,
revealing that the model presents positive-definite energy for small values of
the Lorentz-violating parameters. We work out the complete dispersion
relations of this planar model from the vacuum-vacuum amplitude, showing that
all theory is nonbirefringent. Such planar model provides a more direct way to
analyze consistency aspects associated with the Feynman propagator and the
effects of the LIV parameters on some planar systems of interest.

This work is organized as follows. In Sec. II, we accomplish the dimensional
reduction of Lagrangian (\ref{L2}), obtaining a planar scalar electrodynamics
in which the Lorentz violation is controlled by the symmetric tensor,
$\kappa_{\mu\rho},$ the counterpart of the original tensor $\kappa_{\hat{\mu
}\hat{\rho}}$ defined in (1+2) dimensions, and a three-vector denoted as
$S_{\nu}$. The energy-momentum tensor is computed and the density of energy is
analyzed. The Sec. III is devoted to the analysis of the dispersion relation
in two situations: considering the complete model and regarding the gauge and
scalar sector as decoupled. In Sec.IV, we write the corresponding equations of
motion and wave equations for the model. The wave equations for the gauge and
scalar sectors are solved in the stationary regime at first-order in the LIV
parameters. In Sec. V, we present our Conclusions.

\section{The dimensional reduction procedure}

In this section, we perform the dimensional reduction of the model represented
by Lagrangian (\ref{L2}). There are some distinct procedures for accomplishing
the dimensional reduction of a theory. In the present case, we adopt the one
that freezes the third spacial component of the position four--vector and any
other four-vector. This is done requiring that the physical fields $\left\{
\chi\right\}  $ do not depend anymore on this component, that is,\ $\partial
_{_{3}}\chi=0$. Besides this, we split out the fourth component of the
four-vectors. This procedure is employed in Ref. \cite{DreducCPT}. The
electromagnetic four-potential is written as
\begin{equation}
A^{\hat{\nu}}\longrightarrow(A^{\nu};\phi),
\end{equation}
where $A^{\left(  3\right)  }=\phi$ is now a scalar field and the Greek
indices (without hat) run from $0$ to $2$, $\mu=0,1,2$. Carrying out this
prescription for the terms of \ Lagrangian (\ref{L2}), one then obtains:
\begin{align}
F_{\hat{\mu}\hat{\nu}}F^{\hat{\mu}\hat{\nu}}  & =F_{\mu\nu}F^{\mu\nu
}-2\partial_{\mu}\phi\partial^{\mu}\phi,\label{F1}\\[0.3cm]
\kappa_{\hat{\nu}\hat{\rho}}F_{\hat{\lambda}}{}^{\hat{\nu}}F^{\hat{\lambda
}\hat{\rho}}  & =\kappa_{\nu\rho}F_{\lambda}{}^{\nu}F^{\lambda\rho}-2S_{\nu
}F^{\nu\lambda}\partial_{\lambda}\phi+\eta\partial_{\lambda}\phi
\partial^{\lambda}\phi-\kappa_{\nu\rho}\partial^{\nu}\phi\partial^{\rho}\phi,
\end{align}
where we have defined $F^{\mu3}=\partial^{\mu}\phi,$ \ $F_{\mu3}%
=-\partial_{\mu}\phi$. Also, we have renamed the set of LIV parameters, \ they
now are represented by a second rank tensor $\kappa_{\nu\rho}$ which is the
(1+2)-dimensional counterpart of the tensor $\kappa_{\widehat{\nu
}\widehat{\rho}}$, a vector $S_{\nu}$ and a scalar quantity $\eta$ which are
defined as%
\begin{equation}
S_{\nu}=\kappa_{\nu3}\ ,\text{ }\eta=\kappa_{33},
\end{equation}
respectively. Thus, after the dimensional reduction procedure we attain the
following Lagrangian density:%
\begin{equation}
\mathcal{L}_{1+2}=\underset{\mathcal{L}_{EM}}{\underbrace{-\frac{1}{4}%
F_{\mu\nu}F^{\mu\nu}-\frac{1}{2}\kappa_{\nu\rho}F_{\lambda}^{\text{ \ }\nu
}F^{\lambda\rho}}}+\underset{\mathcal{L}_{scalar}}{\underbrace{\frac{1}%
{2}[1-\eta]\partial_{\mu}\phi\partial^{\mu}\phi+\frac{1}{2}\kappa_{\nu\rho
}\partial^{\nu}\phi\partial^{\rho}\phi}+\underset{\mathcal{L}_{coupling}%
}{\underbrace{S_{\nu}F^{\nu\lambda}\partial_{\lambda}\phi}}}-A_{\mu}J^{\mu
}-J\phi,\label{LR}%
\end{equation}
it is composed of a gauge sector $\left(  \mathcal{L}_{EM}\right)  ,$ a scalar
sector $\left(  \mathcal{L}_{scalar}\right)  ,$ and a coupling sector $\left(
\mathcal{L}_{coupling}\right)  \ $ruled by the Lorentz-violating vector
$S_{\nu}$ that contains three LIV parameters. The Lorentz-violating symmetric
tensor $\kappa_{\nu\rho}$ presents six independent coefficients, which modify
both the electromagnetic and scalar sectors, altering the dynamics of the
Maxwell field and yielding a noncanonical kinetic term for the scalar field.
The LIV noncanonical kinetic term present in the scalar sector has been
recently investigated in scenarios involving topological defects in (1+1)
dimensions \cite{Defects} and acoustic black holes with Lorentz-violation
\cite{Brito} in (1+2) dimensions. A similar term is also found in the
Lagrangian of Ref.\cite{DreducCPT}. The present work provides a possible
origin for this kind of term.

Our planar model (\ref{LR}) has ten dimensionless Lorentz-violating parameters
contained in the tensors $\kappa_{\nu\rho}$, $S_{\nu}$ and in the scalar
$\eta$. The traceless condition of the original tensor, $\kappa^{\hat{\rho}}%
{}_{\hat{\rho}}=0$, gives one constraint between the $\kappa_{\nu\rho}%
-$components
\begin{equation}
\kappa_{00}-\kappa_{ii}=\eta,\label{trace}%
\end{equation}
so, the model possesses nine independent Lorentz-violating parameters, the
same number of the original four-dimensional theory. It demonstrates the
consistency in the dimensional reduction procedure.

We define the components of the electric field as $E^{i}=F_{0i}$, the magnetic
field by $B=-\frac{1}{2}\epsilon_{ij}F_{ij}$ and $\epsilon_{012}=\epsilon
_{12}=1$, then the Lagrangian (\ref{LR}) can be written in terms of fields of
the electric and magnetic field in the form:
\begin{equation}
\mathcal{L}_{1+2}=\mathcal{L}_{EM}+\mathcal{L}_{scalar}+\mathcal{L}%
_{coupling},
\end{equation}
where
\begin{align}
\mathcal{L}_{EM} & =\frac{1}{2}(1+\kappa_{00})\mathbf{E}^{2}-\frac{1}%
{2}(1-\kappa_{ii})B^{2}-\frac{1}{2}\kappa_{ij}E^{i}E^{j}+\kappa_{0i}%
\epsilon_{ij}E^{j}B,\\[0.3cm]
\mathcal{L}_{scalar} & =\frac{1}{2}(1-\eta)[\left(  \partial_{t}\phi\right)
^{2}-\left(  \partial_{i}\phi\right)  ^{2}]+\frac{1}{2}\kappa_{00}\left(
\partial_{t}\phi\right)  ^{2}-\kappa_{0i}\partial_{t}\phi\partial_{i}%
\phi+\frac{1}{2}\kappa_{ij}\partial_{i}\phi\partial_{j}\phi,\\[0.3cm]
\mathcal{L}_{coupling} & =-S^{0}E^{j}\partial_{j}\phi-S^{i}E^{i}\partial
_{t}\phi+\epsilon_{ij}S^{i}\partial_{j}\phi B.\label{cccg}%
\end{align}

The above decomposition allows to determine the parity-properties of the LIV
coefficients. In (1+2)-dimension, the parity operator acts doing
$\mathbf{r\rightarrow}(-x,y)$, it changes the fields as $A_{0}\rightarrow
A_{0},$ $\mathbf{A\rightarrow}(-A_{x},A_{y})$\textbf{, }the
$\mathbf{E\rightarrow}(-E_{x},E_{y})\mathbf{,}$ $B\rightarrow-B$. For more
details, see Ref. \cite{Deser}. Here, we consider that the field $\phi$
behaves as a scalar, $\phi\rightarrow\phi$. Since the Lagrangian density is
parity-even, we can conclude that the planar model possesses nine independent
coefficients, six are parity-even $\left(  \kappa_{00},\kappa_{02},\kappa
_{11},\kappa_{22},S_{0},S_{2}\right)  $, and three are parity-odd\textbf{\ }%
$\left(  \kappa_{01},\kappa_{12},S_{1}\right)  $. The fact that the components
of the vector $S^{\mu}$ transform distinctly is a consequence of the way as
the vectors $\mathbf{E,B}$ $\ $and the field $\phi$ behave under parity.

An issue that deserves some attention is the energy stability, once it is
known that the Lorentz violation yields energy instability in some models, as
for example, the Carroll-Field-Jackiw electrodynamics \cite{Jackiw}. A
preliminary analysis concerning this point can be performed by means of the
energy-momentum tensor for the full planar theory,
\begin{equation}
\Theta^{\mu\nu}=\frac{\partial\mathcal{L}}{\partial\left(  \partial_{\mu
}A_{\rho}\right)  }\partial^{\nu}A_{\rho}+\frac{\partial\mathcal{L}}%
{\partial\left(  \partial_{\mu}\phi\right)  }\partial^{\nu}\phi-g^{\mu\nu
}\mathcal{L},
\end{equation}
which is carried out as%
\begin{align}
\Theta^{\mu\nu}  & =-F^{\mu\rho}F^{\nu}{}_{\rho}-\kappa^{\rho\beta}F^{\mu}%
{}_{\beta}F^{\nu}{}_{\rho}+\kappa^{\mu\beta}F^{\rho}{}_{\beta}F^{\nu}{}_{\rho
}+S^{\mu}F^{\nu\rho}{}\partial_{\rho}\phi\nonumber\\[-0.3cm]
& \\
& +S_{\rho}F^{\rho\nu}{}\partial^{\mu}\phi+S_{\beta}F^{\beta\mu}\partial^{\nu
}\phi+\left(  1-\eta\right)  \partial^{\mu}\phi\partial^{\nu}\phi+\kappa
^{\mu\beta}\partial_{\beta}\phi\partial^{\nu}\phi-g^{\mu\nu}\mathcal{L}%
.\nonumber
\end{align}
We now specialize our evaluation for the density of energy ,%
\begin{equation}
\Theta{^{00}}={\frac{1}{2}M_{jk}E_{j}E_{k}}+{\frac{1}{2}\left(  1-\kappa
_{jj}\right)  B^{2}+B\epsilon_{jk}S_{j}\partial_{k}\phi-S_{j}E_{j}{}%
\partial_{0}}\phi+{\frac{1}{2}\left(  1+\kappa_{jj}\right)  \left(
\partial_{0}\phi\right)  ^{2}\ +\frac{1}{2}N_{jk}\partial_{j}\phi\partial
_{k}\phi,}\label{T00}%
\end{equation}
where we have defined the symmetric matrices%
\begin{equation}
M_{jk}={\left(  1+\kappa_{00}\right)  \delta_{jk}-\kappa_{jk}~\ ,~\ \ N}%
_{jk}={{\left(  1-\kappa_{00}+\kappa_{ii}\right)  \delta}_{jk}-\kappa{_{jk},}}%
\end{equation}
and used $\eta=\kappa_{00}-\kappa_{jj}$. We see that the energy density for
the electromagnetic and scalar fields, when regarded as isolated, are
\begin{align}
\Theta_{EM}^{00}  & ={\frac{1}{2}}M_{ij}{E_{j}E_{k}+\frac{1}{2}\left(
1-\kappa_{jj}\right)  B^{2}},~\label{T00G}\\[0.3cm]
\Theta_{scalar}^{00}  & ={\frac{1}{2}\left(  1+\kappa_{jj}\right)  \left(
\partial_{0}\phi\right)  ^{2}\ +\frac{1}{2}}N_{jk}{\partial_{j}\phi
\partial_{k}\phi}.\label{T00S}%
\end{align}
Both the gauge and scalar energy densities will be positive-definite if
$\left\vert \kappa_{jj}\right\vert <1$ and the matrices $M_{ij}$ and $N_{ij}$
are positive-definite. As the LV parameters are usually much smaller than the
unit, we conclude that the scalar and gauge sectors, as regarded separately,
are stable. However, the energy positivity of the full model seems to be
spoiled by the mixing terms, {$S$}${_{j}E_{j}{}\partial_{0}}\phi\ $and
${B\epsilon_{jk}}${$S$}${_{j}\partial_{k}\phi}$. In order to have more
clarity, we write Eq.(\ref{T00}) in the following way
\begin{align}
\Theta{^{00}}  & ={\frac{1}{2}}\left[  E_{j}-\left(  M^{-1}\right)
_{ja}{S_{a}\partial_{0}\phi}\right]  {M_{jk}}\left[  E_{k}-\left(
M^{-1}\right)  _{ka}{S_{a}\partial_{0}\phi}\right]  +{\frac{1}{2}\left(
1-\kappa_{ii}\right)  }\left[  B+\frac{{\epsilon_{jk}S_{j}\partial_{k}\phi}%
}{{\left(  1-\kappa_{ii}\right)  }}\right]  ^{2}\nonumber\\[-0.3cm]
& \\
& +{\frac{1}{2}\left[  1+\kappa_{jj}-\left(  M^{-1}\right)  _{ij}{S_{i}S}%
_{j}\right]  \left(  \partial_{0}\phi\right)  ^{2}\ +\frac{1}{2}}\left[
{N_{jk}-}\frac{\left(  S_{a}\right)  ^{2}\delta_{jk}-S_{j}S_{k}}%
{{1-\kappa_{ii}}}\right]  {\partial_{j}\phi\partial_{k}\phi.}\nonumber
\end{align}
It shows that the energy density is positive-definite whenever the LV
parameters are sufficiently small.

\section{Dispersion relations}

In this section, we compute the dispersion relations of the model described by
the Lagrangian density (\ref{LR}). Our approach follows an alternative way by
evaluating the vacuum-vacuum amplitude (VVA) of the model. After the
Hamiltonian analysis, the well-defined vacuum-vacuum amplitude (VVA) for the
model, in the generalized Lorentz gauge, can be written as
\begin{equation}
Z=\det\left(  \xi^{-1/2}\square\right)  \int\mathcal{D}A_{\mu}\mathcal{D}%
\phi\exp\left\{  i\int dx~\frac{1}{2}A_{\mu}D^{\mu\nu}A_{\nu}-\frac{1}{2}%
\phi\mathbb{\boxdot}\phi+\phi\mathbb{J}^{\mu}A_{\mu}\right\}  ,\label{zz-1}%
\end{equation}
where $\xi$ is the gauge-fixing parameter and we have defined the following operators:%

\begin{equation}
D^{\mu\nu}=\left(  \square+\kappa^{\rho\sigma}\partial_{\rho}\partial_{\sigma
}\right)  g^{\mu\nu}+\left(  {\xi}^{-1}-1\right)  \partial^{\mu}\partial^{\nu
}+\kappa^{\mu\nu}\square-\kappa^{\mu\rho}\partial_{\rho}\partial^{\nu}%
-\kappa^{\nu\rho}\partial_{\rho}\partial^{\mu},\label{DDg}%
\end{equation}%
\begin{equation}
\boxdot=\left(  1-\eta\right)  \square+\kappa^{\mu\nu}\partial_{\mu}%
\partial_{\nu}~,~\ \mathbb{J}^{\mu}=S^{\mu}\square-S^{\nu}\partial_{\nu
}\partial^{\mu}.\label{FFs}%
\end{equation}

With the purpose of understanding the dispersion relations of the full model,
we first analyze the dispersion relations of the gauge and scalar sectors when
considered uncoupled.

\subsection{Uncoupled dispersion relations}

For $S^{\mu}=0$ the vacuum-vacuum amplitude (\ref{zz-1}) is factored as
$Z=Z_{A_{\mu}}Z_{\phi},$ where $Z_{A_{\mu}}$ and $Z_{\phi}$ are the
vacuum-vacuum amplitudes for the pure gauge and pure scalar fields, respectively.

\subsubsection{Dispersion relation for the pure gauge field}

The vacuum-vacuum amplitude for the pure gauge field is
\begin{equation}
Z_{A_{\mu}}=\det\left(  \xi^{-1/2}\square\right)  \int\mathcal{D}A_{\mu}%
\exp\left\{  i\int dx~\frac{1}{2}A_{\mu}D^{\mu\nu}A_{\nu}\right\}
=\det\left(  \xi^{-1/2}\square\right)  \left(  \det D^{\mu\nu}\right)
^{-1/2},
\end{equation}
with the operator $D^{\mu\nu}$ defined by (\ref{DDg}). By computing the
functional determinant,
\begin{equation}
\det D^{\mu\nu}=\det\left(  \xi^{-1}\square^{2}\right)  \det\left(
\mathbb{\boxminus}\right)  ,
\end{equation}
the VVA for the pure gauge field is
\begin{equation}
Z_{A_{\mu}}=\det\left(  \mathbb{\boxminus}\right)  ^{-1/2},
\end{equation}
where the operator $\mathbb{\boxminus}$ in momentum space reads as
\begin{equation}
\mathbb{\tilde{\boxminus}}=\alpha p_{0}^{2}+\beta p_{0}+\gamma,
\end{equation}
\textbf{with }the coefficients defined as%
\begin{align}
\alpha & =(1+\kappa_{00})(1+\kappa_{00}-\text{tr}\mathbb{K)}+\det
\mathbb{K},~\ \ \mathbb{K}=\left[  \kappa_{ij}\right]
,\ \ \ \ \ \ \ \\[0.3cm]
\text{ }\beta & =-2\kappa_{0i}Q_{ij}p_{j},\text{ \ }Q_{ij}\;=\;\left[
(1+\kappa_{00})\delta_{ij}-\kappa_{ij}\right]  ,\\[0.3cm]
\gamma & =(1-\text{tr}\mathbb{K})[\kappa_{ij}p_{i}p_{j}-(1+\kappa
_{00})\mathbf{p}^{2}]-(\epsilon_{ij}p_{i}\kappa_{0j})^{2}.
\end{align}

The dispersion relations for the pure gauge field are obtained from the
condition $\mathbb{\tilde{\boxminus}}=0$, which yields
\begin{equation}
p_{0}=\frac{\kappa_{0i}Q_{ij}p_{j}}{\alpha}\pm\frac{\sqrt{(\kappa_{0i}%
Q_{ij}p_{j})^{2}-\alpha(1-\text{tr}\mathbb{K})[\kappa_{ij}p_{i}p_{j}%
-(1+\kappa_{00})\mathbf{p}^{2}]+\alpha(\epsilon_{ij}p_{i}\kappa_{0j})^{2}}%
}{\alpha}.\label{DR2}%
\end{equation}
It is easy to show that this relation implies nonbirefringence at any order in
the LIV parameters, once it yields the same phase velocity for the left and
right modes traveling at the same sense. For similar situations, see Ref.
\cite{DreducCPT}. At first order, it is given by
\begin{equation}
p_{0}=\kappa_{0i}p_{i}\pm\left\vert \mathbf{p}\right\vert \left(  1-\frac
{1}{2}\kappa_{00}-\frac{\kappa_{ij}p_{i}p_{j}}{2\mathbf{p}^{2}}\right)
.\label{FO1}%
\end{equation}

The gauge dispersion relation (\ref{DR2}) can specialized for some particular
cases. For $\kappa_{ij}=0,$ $\kappa_{0j}=0,$ the Lorentz-violating
coefficients are represented by the parity-even element $\kappa_{00}$ and the
Eq.(\ref{DR2}) yields the relation%
\begin{equation}
p_{0}=\pm\frac{\left\vert \mathbf{p}\right\vert }{(1+\kappa_{00})^{1/2}%
},\label{DR2A}%
\end{equation}
which is the isotropic parity-even dispersion relation. Adopting $\kappa
_{00}=0,\kappa_{0j}=0,$ we achieve the anisotropic dispersion relation,%
\begin{equation}
p_{0}=\pm N_{0}\left\vert \mathbf{p}\right\vert \sqrt{1-\kappa_{ij}p_{i}%
p_{j}/\mathbf{p}^{2}},\label{DR2B}%
\end{equation}
where $N_{0}=\sqrt{(1-\text{tr}\mathbb{K})/(1-\text{tr}\mathbb{K}%
+\det\mathbb{K})}$. This relation involves parity-even and parity-odd coefficients.

For $\kappa_{ij}=0,$ $\kappa_{00}=0,$ we attain other anisotropic dispersion
relation,%
\begin{equation}
p_{0}=\kappa_{0i}p_{i}\pm\left\vert \mathbf{p}\right\vert \sqrt{1+\left(
\kappa_{0i}\right)  ^{2}}.\label{DR2C}%
\end{equation}

The energy-momentum tensor of the pure gauge field shows that the
electromagnetic sector represents a stable theory. The relations
(\ref{DR2A},\ref{DR2B},\ref{DR2C}), however, could anticipate a noncausal
electrodynamics for some values of the LIV coefficients. The spoil of
causality may be inferred from the evaluation of the group velocity $\left(
u_{g}=dp_{0}/d\left\vert \mathbf{p}\right\vert \right)  $ associated with each
dispersion relation.

\subsubsection{Dispersion relation of the pure scalar sector}

In the same way, the vacuum-vacuum amplitude for the uncoupled scalar field
is
\begin{equation}
Z_{\phi}=\int\mathcal{D}\phi\exp\left\{  -\frac{i}{2}\int dx\phi
\mathbb{\boxdot}\phi\right\}  =\left(  \det\mathbb{\boxdot}\right)  ^{-1/2},
\end{equation}
with the operator $\mathbb{\boxdot}$ defined in Eq. (\ref{FFs}). In the
momentum space it is read as
\begin{equation}
\mathbb{\tilde{\boxdot}}=\left(  1-\eta\right)  p^{2}+\tilde{\kappa}%
^{\rho\sigma}p_{\rho}p_{\sigma}.\
\end{equation}
The dispersion relation of the scalar field are computed by the condition
$\mathbb{\tilde{\boxdot}}=0,$ taking into account the relation (\ref{trace}),
which provides the following equation for $p_{0}$:
\begin{equation}
\left(  1+\text{tr}\mathbb{K}\right)  p_{0}^{2}-2\left(  \kappa_{0i}%
p_{i}\right)  p_{0}-\left(  1-\kappa_{00}+\text{tr}\mathbb{K}\right)
\mathbf{p}^{2}+\kappa_{ij}p_{i}p_{j}=0,
\end{equation}
whose roots are
\begin{equation}
p_{0}^{\left(  \pm\right)  }=\lambda\left[  \kappa_{0i}p_{i}\pm\sqrt{\left(
\kappa_{0i}p_{i}\right)  ^{2}+\left(  1+\text{tr}\mathbb{K}\right)  \left[
\left(  1-\kappa_{00}+\text{tr}\mathbb{K}\right)  \mathbf{p}^{2}-\kappa
_{ij}p_{i}p_{j}\right]  }\right]  ,\label{DRs}%
\end{equation}
where $\lambda=\left[  1+\text{tr}\mathbb{K}\right]  ^{-1}.$ This is a
nonbirefringent relation at any order in LIV parameters. At first order such
relation is given by
\begin{equation}
p_{0}=\kappa_{0i}p_{i}\pm\left\vert \mathbf{p}\right\vert \left(  1-\frac
{1}{2}\kappa_{00}-\frac{\kappa_{ij}p_{i}p_{j}}{2\mathbf{p}^{2}}\right)
,\label{FO2}%
\end{equation}
which is exactly the first-order gauge dispersion relation given in Eq.
(\ref{FO1}). Although the exact dispersion relations of the scalar and gauge
sectors, (\ref{DR2}) and (\ref{DRs}), are clearly different, at first order in
the LIV parameters both sectors are governed by the same dispersion relation.
A direct analysis of the relation (\ref{DRs}) indicates that the scalar sector
can support noncausal modes, similarly as it occurs in the gauge sector.

\subsection{Full dispersion relations}

In order to examine the complete dispersion relations, we evaluate the
vacuum-vacuum amplitude (\ref{zz-1}) considering the presence of the coupling
vector, $S^{\mu}$. We first integrate the $\phi-$field, obtaining
\begin{equation}
Z=\det\left(  \xi^{-1/2}\square\right)  \det\left(  \mathbb{\boxdot}\right)
^{-1/2}\int\mathcal{D}A_{\mu}\exp\left\{  i\int dx~\frac{1}{2}A_{\mu
}\mathbb{D}^{\mu\nu}A_{\nu}\right\}  ,
\end{equation}
where the operator $\mathbb{D}^{\mu\nu}$ is defined as%
\begin{equation}
\mathbb{D}^{\mu\nu}=D^{\mu\nu}+\frac{\mathbb{J^{\mu}J}^{\nu}}{\mathbb{\tilde
{\boxdot}}}.
\end{equation}
By integrating the gauge field, we achieve%
\begin{equation}
Z=\det\left(  \xi^{-1/2}\square\right)  \det\left(  \mathbb{\boxdot}\right)
^{-1/2}\det\left(  \mathbb{D}^{\mu\nu}\right)  ^{-1/2},
\end{equation}
which can be rewritten as
\begin{equation}
Z=\det\left(  \xi^{-1/2}\square\right)  \det\left(  \mathbb{\boxdot}\right)
\det\left(  \mathbb{\boxdot}D^{\mu\nu}+\mathbb{J^{\mu}J}^{\nu}\right)
^{-1/2}.\label{ssd}%
\end{equation}
We now compute the functional determinant of the term $\left(  \mathbb{\boxdot
}D^{\mu\nu}+\mathbb{J^{\mu}J}^{\nu}\right)  $,
\begin{equation}
\det\left(  \mathbb{\boxdot}D^{\mu\nu}+\mathbb{J^{\mu}J}^{\nu}\right)
=\det\left(  \xi^{-1/2}\square\right)  ^{2}\det\left(  \mathbb{\boxdot
}\right)  ^{2}\det\left(  \otimes\right)  ,
\end{equation}
which replaced in Eq. (\ref{ssd}) leads to the simpler result%
\begin{equation}
Z=\det\left(  \otimes\right)  ^{-1/2}.\label{zzx}%
\end{equation}

In the momentum space the operator $\otimes$ is represented by $\tilde
{\otimes}\left(  p\right)  $ and the dispersion relations for the full model
are obtained from the equation $\tilde{\otimes}\left(  p\right)  =0$. In our
case, we have the exact equation for the dispersion relations,\textbf{\ }%
\begin{equation}
\tilde{\otimes}\left(  p\right)  =a_{4}\left(  p_{0}\right)  ^{4}+a_{3}\left(
p_{0}\right)  ^{3}+a_{2}\left(  p_{0}\right)  ^{2}+a_{1}p_{0}+a_{0}%
=0,\label{fulldp-1}%
\end{equation}
with $a_{k}$ $\left(  k=0,1,2,3,4,\right)  $ being functions of the LIV
parameters having the following structure%
\begin{align}
a_{4}  & =1+a_{4}^{\left(  1\right)  }+a_{4}^{\left(  2\right)  }%
+a_{4}^{\left(  3\right)  },~\ \\[0.2cm]
a_{3}  & =a_{3}^{\left(  1\right)  }+a_{3}^{\left(  2\right)  }+a_{3}^{\left(
3\right)  },\\[0.2cm]
a_{2}  & =-2\mathbf{p}^{2}+a_{2}^{\left(  1\right)  }+a_{2}^{\left(  2\right)
}+a_{2}^{\left(  3\right)  },~\\[0.2cm]
a_{1}  & =a_{1}^{\left(  1\right)  }+a_{1}^{\left(  2\right)  }+a_{1}^{\left(
3\right)  },\\[0.2cm]
a_{0}  & =\mathbf{p}^{4}+a_{0}^{\left(  1\right)  }+a_{0}^{\left(  2\right)
}+a_{0}^{\left(  3\right)  },
\end{align}
where $a_{k}^{\left(  n\right)  }$ $\left(  n=1,2,3\right)  $ represents the
contribution to $n$th order in the LIV parameters to the coefficient $a_{k}$,
whose explicit expressions are given in the appendix A. Below we present some
configurations of the LIV parameters which allow to factorize and solve
exactly the full dispersion relation equation given in Eq\textbf{.}%
(\ref{fulldp-1}).

We first analyze the pure contribution of the coupling vector $S^{\mu}$ to the
dispersion relations of the scalar and gauge fields. For this purpose, we set
$\kappa^{\mu\nu}=0$ in the full vacuum-vacuum amplitude (\ref{zzx}), obtaining%
\begin{equation}
Z=\det\left(  \square\right)  ^{-1/2}\det\left[  \left(  1+S^{2}\right)
\square-\left(  S\cdot\partial\right)  ^{2}\right]  ^{-1/2}.
\end{equation}
It describes two bosonic degrees of freedom; a first one is a gauge field
governed by the usual dispersion relation,
\begin{equation}
p_{0}=\pm\left\vert \mathbf{p}\right\vert ,
\end{equation}
while the second one describes a massless scalar field
\begin{equation}
\left(  p_{0}\right)  _{\pm}=-\frac{S_{0}\left(  \mathbf{S\cdot p}\right)
}{1-\mathbf{S}^{2}}\pm\frac{\sqrt{\mathbf{p}^{2}\left(  1+S^{2}\right)
\left(  1-\mathbf{S}^{2}\right)  +\left(  \mathbf{S\cdot p}\right)
^{2}\left(  1-\mathbf{S}^{2}\right)  +\left(  S_{0}\right)  ^{2}\left(
\mathbf{S\cdot p}\right)  ^{2}}}{1-\mathbf{S}^{2}},
\end{equation}
which also is compatible with absence of birefringence. At leading-order the
above dispersion relation reads as
\begin{equation}
\left(  p_{0}\right)  _{\pm}=-S_{0}\left(  \mathbf{S\cdot p}\right)
\pm\left\vert \mathbf{p}\right\vert \left(  1+\frac{1}{2}\left(  S_{0}\right)
^{2}+\frac{1}{2}\frac{\left(  \mathbf{S\cdot p}\right)  }{\mathbf{p}^{2}}%
^{2}\right)  ,
\end{equation}
showing that the contributions of the vector $S_{\mu}$ to the dispersion
relations only begin at second order.

The second case corresponds to the general isotropic dispersion relation,
provided by fixing $\kappa_{ij}=0$, $\kappa_{0i}=0$ and $S_{i}=0$. The
partition function (\ref{zzx}) factorizes as
\begin{equation}
Z=\det\left[  \left(  1+\kappa_{00}\right)  \square+\kappa_{00}\nabla
^{2}\right]  ^{-1/2}\det\left[  \left(  1+\kappa_{00}\right)  \square-\left\{
\left(  S_{0}\right)  ^{2}-\left(  k_{00}\right)  ^{2}-\kappa_{00}\right\}
\nabla^{2}\right]  ^{-1/2},
\end{equation}
describing two bosonic degree of freedom supporting the following dispersion
relations:
\begin{equation}
p_{0}=\pm\frac{\left\vert \mathbf{p}\right\vert }{\sqrt{1+\kappa_{00}}%
},\label{G11}%
\end{equation}
\begin{equation}
p_{0}=\pm\left\vert \mathbf{p}\right\vert \sqrt{\frac{1-\left(  \kappa
_{00}\right)  ^{2}+\left(  S_{0}\right)  ^{2}}{1+\kappa_{00}}}.\label{S11}%
\end{equation}
The relation (\ref{G11}) describes the gauge field, while the relation
(\ref{S11}) is associated to the massless scalar field. This association comes
from Eqs. (\ref{DR2}) and (\ref{DRs}), when properly written for the pure
isotropic coefficient, $\kappa_{00}.$

A third case is obtained by considering $\kappa_{0i}$ and $S_{0}$ as non-null,
which provides the following vacuum-vacuum amplitude%
\begin{equation}
Z=\det\left[  \square-2\kappa_{0i}\partial_{i}\partial_{0}\right]  ^{-1/2}%
\det\left[  \square-2\kappa_{0i}\partial_{i}\partial_{0}-\left\{  \left(
S_{0}\right)  ^{2}+\left(  k_{0i}\right)  ^{2}\right\}  \nabla^{2}\right]
^{-1/2}.
\end{equation}
The first operator $\square-2\kappa_{0i}\partial_{i}\partial_{0}$ describes
the dispersion relation of a massless scalar degree of freedom%
\begin{equation}
p_{0}=\kappa_{0i}p_{i}\pm\left\vert \mathbf{p}\right\vert \sqrt{1+\frac
{\left(  \kappa_{0i}p_{i}\right)  ^{2}}{\mathbf{p}^{2}}},
\end{equation}
while the operator $\square-2\kappa_{0i}\partial_{i}\partial_{0}-\left[
\left(  S_{0}\right)  ^{2}+\left(  k_{0i}\right)  ^{2}\right]  \nabla^{2}$
gives the dispersion relations of the gauge field,
\begin{equation}
p_{0}=\kappa_{0i}p_{i}\pm\left\vert \mathbf{p}\right\vert \sqrt{1+\left(
\kappa_{0i}\right)  ^{2}+\left(  S_{0}\right)  ^{2}}.
\end{equation}
The specialization of the exact relations (\ref{DR2}) and (\ref{DRs}) for the
coefficients $\kappa_{0i}$ is the element that allows to define what is the
scalar and the gauge field dispersion relation.

A more complicate case which also provides exact dispersion relations is
obtained by considering as non-null $\kappa_{00}$ and $S_{i}$, yielding
\begin{equation}
\tilde{\otimes}\left(  p\right)  =a_{4}\left(  p_{0}\right)  ^{4}-a_{2}\left(
p_{0}\right)  ^{2}+a_{0}=0,
\end{equation}
with%
\begin{align}
a_{4}  & =\left(  1+\kappa_{00}\right)  \left(  1+\kappa_{00}-\mathbf{S}%
^{2}\right)  ,\\
a_{2}  & =\mathbf{p}^{2}\left(  1+\kappa_{00}\right)  \left[  2-\left(
\kappa_{00}\right)  ^{2}-2\mathbf{S}^{2}\right]  +\left(  1+2\kappa
_{00}\right)  \left(  \mathbf{S\cdot p}\right)  ^{2},\\
a_{0}  & =\mathbf{p}^{4}\left(  1+\kappa_{00}\right)  \left(  1-\kappa
_{00}-\mathbf{S}^{2}\right)  +\mathbf{p}^{2}\left(  1+\kappa_{00}\right)
\left(  \mathbf{S\cdot p}\right)  ^{2}.
\end{align}
It gives the dispersion relation for the gauge field
\begin{equation}
p_{0}^{\left(  1\right)  }=\pm\sqrt{\frac{a_{2}+\sqrt{\left(  a_{2}\right)
^{2}-4a_{4}a_{0}}}{2a_{4}}},
\end{equation}
and the following one for the massless scalar
\begin{equation}
p_{0}^{\left(  2\right)  }=\pm\sqrt{\frac{a_{2}-\sqrt{\left(  a_{2}\right)
^{2}-4a_{4}a_{0}}}{2a_{4}}}.
\end{equation}

Both dispersion relations can expressed at second order in the LIV
coefficients, yielding
\begin{align}
p_{0}  & =\pm\left\vert \mathbf{p}\right\vert \left(  1-\frac{1}{2}\kappa
_{00}+\frac{A^{\left(  2\right)  }-2\sqrt{B^{\left(  4\right)  }}}%
{8\mathbf{p}^{2}}\right)  ,\label{dpx1}\\
p_{0}  & =\pm\left\vert \mathbf{p}\right\vert \left(  1-\frac{1}{2}\kappa
_{00}+\frac{A^{\left(  2\right)  }+2\sqrt{B^{\left(  4\right)  }}}%
{8\mathbf{p}^{2}}\right)  ,\label{dpx2}%
\end{align}
where%
\begin{align}
A^{\left(  2\right)  }  & =\mathbf{p}^{2}\left(  \kappa_{00}\right)
^{2}+2\left(  \mathbf{S\cdot p}\right)  ^{2},\\[0.2cm]
B^{\left(  4\right)  }  & =\mathbf{p}^{4}\left(  \kappa_{00}\right)
^{4}+4\mathbf{p}^{4}\left(  \kappa_{00}\right)  ^{2}\mathbf{S}^{2}%
-6\mathbf{p}^{2}\left(  \kappa_{00}\right)  ^{2}\left(  \mathbf{S\cdot
p}\right)  ^{2}+\left(  \mathbf{S\cdot p}\right)  ^{4}.
\end{align}
Here, it is important to highlight that at first-order in the LIV backgrounds
the dispersion relations (\ref{dpx1},\ref{dpx2}) are the same one, confirming
the results of the previous subsections: at first order the scalar and the
gauge sectors are governed by the same dispersion relations.

For arbitrary configurations of the LIV\ backgrounds, it is convenient to
compute the roots of the dispersion relations (\ref{fulldp-1}) in a
perturbative way. At first order in the LIV parameters, we obtain
\begin{equation}
p_{0}^{\left(  g,s\right)  }=\kappa_{0i}p_{i}\pm\left\vert \mathbf{p}%
\right\vert \left(  1-\frac{1}{2}\kappa_{00}-\frac{1}{2}\frac{\kappa_{ij}%
p_{i}p_{j}}{\mathbf{p}^{2}}\right)  ,\label{ggff}%
\end{equation}
for the dispersion relations of the gauge and massless scalar fields. This is
the same expression of Eqs. (\ref{FO1}, \ref{FO2}), confirming our previous
computations. We thus verify that the all the dispersion relations of this
planar model are free from the influence of the vector $S^{\mu}$ at
first-order in the LIV parameters.

\section{Equations of motion and stationary solutions}

The classical behavior of this theory is governed by the equations of motion
stemming from the Euler-Lagrange equations, that is
\begin{align}
\partial_{\alpha}F^{\alpha\beta}+\kappa^{\beta\rho}\partial_{\alpha}F^{\alpha
}{}_{\rho}-\kappa_{\text{ \ }}^{\alpha\rho}\partial_{\alpha}F^{\beta}{}_{\rho
}+S^{\beta}\square\phi-S^{\alpha}\partial_{\alpha}\partial^{\beta}\phi &
=J^{\beta},\label{G1}\\[0.3cm]
\lbrack1-\eta]\square\phi+\kappa^{\alpha\rho}\partial_{\alpha}\partial_{\rho
}\phi+S_{\nu}\partial_{\alpha}F^{\nu\alpha}  &  =-J.\label{Phi2}%
\end{align}
In terms of the gauge potential and by using the Lorentz gauge, $\partial\cdot
A=0$, these equations are written as%
\begin{align}
\lbrack\square g^{\beta\rho}+\square\kappa_{\text{ }}^{\beta\rho}+g^{\beta
\rho}\kappa_{\text{ \ }}^{\alpha\sigma}\partial_{\alpha}\partial_{\sigma
}-\kappa_{\text{ \ }}^{\rho\alpha}\partial_{\alpha}\partial^{\beta}]A_{\rho
}+\left[  S^{\beta}\square-S^{\alpha}\partial_{\alpha}\partial^{\beta}\right]
\phi &  =J^{\beta},\label{G1B}\\[0.3cm]
\lbrack(1-\eta)\square+\kappa^{\alpha\rho}\partial_{\alpha}\partial_{\rho
}]\phi+S_{\nu}\square A^{\nu}  &  =-J.\label{Phi1B}%
\end{align}

The modified Maxwell equations stems from Eq.(\ref{G1}) lead to the altered
forms for the Gauss's and Ampere's laws,
\begin{equation}
\left(  1+\kappa_{00}\right)  \partial_{i}E^{i}+\epsilon^{ji}\kappa
_{0j}\partial_{i}B-\kappa_{ij}\partial_{i}E^{j}-S_{0}\nabla^{2}\phi
-S^{i}\partial_{i}\partial_{t}\phi\;=\;\rho,\label{E1}%
\end{equation}%
\begin{align}
\left(  \epsilon^{ij}-\kappa_{il}\epsilon^{lj}-\kappa_{jl}\epsilon
^{il}\right)  \partial_{j}B+\kappa_{0l}\epsilon^{il}\partial_{0}B-\partial
_{0}E^{i}+\kappa_{il}\partial_{0}E^{l}-\kappa_{i0}\partial_{j}E^{j}  &
\nonumber\\
& \label{B1}\\
+\kappa_{j0}\partial_{j}E^{i}-S^{i}\nabla^{2}\phi+S^{i}\partial_{0}^{2}%
\phi-S^{j}\partial_{j}\partial^{i}\phi-S^{0}\partial_{0}\partial^{i}\phi &
=J^{i},\nonumber
\end{align}
while the scalar sector evolves in accordance with
\begin{equation}
\lbrack1-\eta+\kappa_{00}]\partial_{t}^{2}\phi-[1-\eta]\nabla^{2}\phi
+\kappa^{ij}\partial_{i}\partial_{j}\phi+2\kappa^{0j}\partial_{0}\partial
_{j}\phi-S_{0}\partial_{i}E^{i}+S_{i}\partial_{0}E^{i}-\epsilon_{ij}%
S_{i}\partial_{j}B=-J.\label{Scalar1}%
\end{equation}

In order to solve this electrodynamics, Eqs.(\ref{Phi2}, \ref{E1}, \ref{B1})
should be considered jointly with the Faraday's law,%
\begin{equation}
\partial_{t}B+\nabla\times\mathbf{E}=0.\label{Bi}%
\end{equation}
which comes from the tensor form of Bianchi identity, $\partial_{\mu}%
F^{\mu\ast}=0.$ Here, $F^{\mu\ast}=\frac{1}{2}\epsilon^{\mu\nu\alpha}%
F_{\nu\alpha}$ is the the dual of the electromagnetic field tensor in $(1+2)-$dimensions.

At first order in LIV parameters, the solutions of the equations of motion
(\ref{G1B}) and (\ref{Phi1B}) are
\begin{align}
A_{\mu}  & =\frac{1}{\square}\left(  g_{\mu\rho}-\kappa_{\mu\rho}-g_{\mu\rho
}\kappa_{\text{ \ }}^{\alpha\beta}\frac{\partial_{\alpha}\partial_{\beta}%
}{\square}+\kappa_{\rho\alpha}\frac{\partial^{\alpha}\partial_{\mu}}{\square
}\right)  J^{\rho}+\frac{1}{\square}\left(  S_{\mu}-S^{\sigma}\frac
{\partial_{\sigma}\partial_{\mu}}{\square}\right)  J,\label{EQGG}\\[0.3cm]
\phi & =-\frac{1}{\square}\left[  1+\eta-\kappa^{\alpha\beta}\frac
{\partial_{\alpha}\partial_{\beta}}{\square}\right]  J+\frac{1}{\square
}S_{\rho}J^{\rho}.\label{EQSC}%
\end{align}
The pure Green's functions for the gauge and the scalar fields read
\begin{align}
G_{\mu\rho}\left(  x-x^{\prime}\right)   & =\frac{1}{\square}\left[
g_{\mu\rho}-\kappa_{\mu\rho}-g_{\mu\rho}\kappa_{\text{ \ }}^{\alpha\beta}%
\frac{\partial_{\alpha}\partial_{\beta}}{\square}+\kappa_{\rho\alpha}%
\frac{\partial^{\alpha}\partial_{\mu}}{\square}\right]  \delta\left(
x-x^{\prime}\right)  ,\label{ggf}\\[0.2cm]
G_{\mu}\left(  x-x^{\prime}\right)   & =\frac{1}{\square}\left(  S_{\mu
}-S^{\sigma}\frac{\partial_{\sigma}\partial_{\mu}}{\square}\right)
\delta\left(  x-x^{\prime}\right)  ,\\[0.2cm]
G\left(  x-x^{\prime}\right)   & =-\frac{1}{\square}\left[  1+\eta-\kappa
^{\mu\beta}\frac{\partial_{\mu}\partial_{\beta}}{\square}\right]
\delta\left(  x-x^{\prime}\right)  ,\label{gsf}%
\end{align}
respectively, where $x=(x_{0},\mathbf{r})$. The above equations show the both
sources $J^{\mu}$ and $J$ can be generate electromagnetic phenomena.

\subsection{Static solutions for the pure gauge field}

The stationary solution for the gauge field in (\ref{EQGG}) can be expressed
as
\begin{equation}
A_{\mu}\left(  \mathbf{r}\right)  =\int d\mathbf{r}^{\prime}G_{\mu\rho}\left(
\mathbf{r-r}^{\prime}\right)  J^{\rho}\left(  \mathbf{r}^{\prime}\right)
+\int d\mathbf{r}^{\prime}G_{\mu}\left(  \mathbf{r-r}^{\prime}\right)
J\left(  \mathbf{r}^{\prime}\right)  ,\label{A5}%
\end{equation}
where $G_{\mu\rho}\left(  \mathbf{r-r}^{\prime}\right)  $ is the stationary
Green's function whose components obtained from (\ref{ggf})\ are
\begin{align}
G_{00}\left(  \mathbf{R}\right)   & =-\frac{1}{2\pi}\left(  1-\kappa
_{00}+\frac{1}{2}\kappa_{aa}\right)  \ln R-\frac{1}{4\pi}\kappa^{ab}%
\frac{R_{a}R_{b}}{R^{2}},\nonumber\\[0.2cm]
G_{0i}\left(  \mathbf{R}\right)   & =\frac{1}{2\pi}\kappa_{0i}\ln
R\;,\;\;G_{i0}\left(  \mathbf{R}\right)  \;=\;\frac{1}{4\pi}\kappa_{0i}\ln
R\text{\ }-\frac{1}{4\pi}\kappa_{0a}\frac{R_{a}R_{i}}{R^{2}},\\[0.2cm]
G_{ij}\left(  \mathbf{R}\right)   & =\frac{1}{2\pi}\left[  \delta_{ij}\left(
1+\frac{1}{2}\kappa_{aa}\right)  +\frac{1}{2}\kappa_{ij}\right]  \ln
R+\frac{1}{4\pi}\delta_{ij}\kappa_{ab}\frac{R_{a}R_{b}}{R^{2}}-\frac{1}{4\pi
}\kappa_{ja}\frac{R_{a}R_{i}}{R^{2}},\nonumber
\end{align}
and $G_{\mu}\left(  \mathbf{r-r}^{\prime}\right)  $ is the Green's functions
describing the contribution of the scalar source $J$ \ to the electromagnetic
field given by%
\begin{equation}
G_{0}\left(  \mathbf{R}\right)  =-\frac{1}{2\pi}S_{0}\ln R~,~\ \ G_{i}\left(
\mathbf{R}\right)  =-\frac{1}{4\pi}S_{i}\ln R\text{\ }+\frac{1}{4\pi}%
S_{a}\frac{R_{a}R_{i}}{R^{2}},
\end{equation}
where we have denoted $\mathbf{R}=\mathbf{r-r}^{\prime}$.

The non-diagonal Green's function components reveal that charges yield
electric and magnetic fields, as well as currents do. We now compute the
electric and magnetic fields for some special configurations of charge and
current densities. In accordance with Eq. (\ref{A5}), the scalar and vector
potentials are
\begin{align}
A_{0}\left(  \mathbf{r}\right)   & =-\frac{1}{2\pi}\left(  1-\kappa_{00}%
+\frac{1}{2}\kappa_{aa}\right)  \int d\mathbf{r}^{\prime}~\rho\left(
\mathbf{r}^{\prime}\right)  ~\ln\left\vert \mathbf{r-r}^{\prime}\right\vert
~-\frac{1}{4\pi}\kappa_{ab}\int d\mathbf{r}^{\prime}\frac{\left(
\mathbf{r-r}^{\prime}\right)  _{a}\left(  \mathbf{r-r}^{\prime}\right)  _{b}%
}{\left(  \mathbf{r-r}^{\prime}\right)  ^{2}}\rho\left(  \mathbf{r}^{\prime
}\right) \nonumber\\[-0.3cm]
& \label{AI}\\
& +\frac{1}{2\pi}\kappa_{0a}\int d\mathbf{r}^{\prime}~J^{a}\left(
\mathbf{r}^{\prime}\right)  \ln\left\vert \mathbf{r-r}^{\prime}\right\vert
~-\frac{1}{2\pi}S_{0}\int d\mathbf{r}^{\prime}J\left(  \mathbf{r}^{\prime
}\right)  \ln\left\vert \mathbf{r-r}^{\prime}\right\vert ~\nonumber
\end{align}
and%
\begin{align}
A_{j}  & =\frac{1}{4\pi}\kappa_{0j}\int d\mathbf{r}^{\prime}~\rho\left(
\mathbf{r}^{\prime}\right)  ~\ln\left\vert \mathbf{r-r}^{\prime}\right\vert
\text{\ }-\frac{1}{4\pi}\kappa_{0a}\int d\mathbf{r}^{\prime}\frac{\left(
\mathbf{r-r}^{\prime}\right)  _{a}\left(  \mathbf{r-r}^{\prime}\right)  _{j}%
}{\left\vert \mathbf{r-r}^{\prime}\right\vert ^{2}}\rho\left(  \mathbf{r}%
^{\prime}\right) \nonumber\\[0.3cm]
& +\frac{1}{2\pi}\left[  \delta_{jb}\left(  1+\frac{1}{2}\kappa_{aa}\right)
+\frac{1}{2}\kappa_{jb}\right]  \int d\mathbf{r}^{\prime}~J^{b}\left(
\mathbf{r}^{\prime}\right)  \ln\left\vert \mathbf{r-r}^{\prime}\right\vert
~\label{AII}\\[0.3cm]
& +\frac{1}{4\pi}\delta_{jc}\kappa_{ab}\int d\mathbf{r}^{\prime}\frac{\left(
\mathbf{r-r}^{\prime}\right)  _{a}\left(  \mathbf{r-r}^{\prime}\right)  _{b}%
}{\left\vert \mathbf{r-r}^{\prime}\right\vert ^{2}}J^{c}\left(  \mathbf{r}%
^{\prime}\right)  -\frac{1}{4\pi}\kappa_{ab}\int d\mathbf{r}^{\prime}%
\frac{\left(  \mathbf{r-r}^{\prime}\right)  _{a}\left(  \mathbf{r-r}^{\prime
}\right)  _{j}}{\left\vert \mathbf{r-r}^{\prime}\right\vert ^{2}}J^{b}\left(
\mathbf{r}^{\prime}\right) \nonumber\\[0.3cm]
& -\frac{1}{4\pi}S_{j}\int d\mathbf{r}^{\prime}J\left(  \mathbf{r}^{\prime
}\right)  \ln\left\vert \mathbf{r-r}^{\prime}\right\vert \text{\ }+\frac
{1}{4\pi}S_{a}\int d\mathbf{r}^{\prime}\frac{\left(  \mathbf{r-r}^{\prime
}\right)  _{a}\left(  \mathbf{r-r}^{\prime}\right)  _{j}}{\left\vert
\mathbf{r-r}^{\prime}\right\vert ^{2}}J\left(  \mathbf{r}^{\prime}\right)
,\nonumber
\end{align}
respectively.

For a pointlike static charge distribution, $\rho(\mathbf{r}^{\prime}%
)=q\delta(\mathbf{r}^{\prime})~\left[  J_{i}\left(  \mathbf{r}^{\prime
}\right)  =0=J\left(  \mathbf{r}^{\prime}\right)  \right]  ,$ the scalar
potential and the potential vector are
\begin{align}
A_{0}\left(  \mathbf{r}\right)   & =-\frac{q}{2\pi}\left[  \left(
1-\kappa_{00}+\frac{1}{2}\kappa_{aa}\right)  \ln r+\frac{1}{2}\kappa_{ab}%
\frac{r_{a}r_{b}}{r^{2}}\right]  ,\label{A01}\\[0.3cm]
A_{j}\left(  \mathbf{r}\right)   & =\frac{q}{4\pi}\left(  \kappa_{0j}\ln
r\text{\ }-\kappa_{0a}\frac{r_{a}r_{j}}{r^{2}}\right)  ,\label{A02}%
\end{align}
respectively. The solution (\ref{A01}) differs from the usual scalar potential
generated by a pointlike charge in (1+2) dimensions mainly by the term
$\kappa^{ab}r_{a}r_{b}/r^{2},$ which yields an anisotropic behavior for it.
The electric field produced by the pointlike charge is,
\begin{equation}
E_{i}\left(  \mathbf{r}\right)  =-\frac{q}{2\pi}\left[  \left(  1-\kappa
_{00}+\frac{1}{2}\kappa_{aa}\right)  \frac{r_{i}}{r^{2}}+\kappa_{ib}%
\frac{r_{b}}{r^{2}}-\kappa_{ab}\frac{r_{a}r_{b}}{r^{4}}r_{i}\right]
,\label{E4}%
\end{equation}
which in addition to its radial behavior $r^{-1}$ presents anisotropies, due
to the two last terms $\kappa_{ib}r_{b}/r^{2}\ $and $\kappa_{ab}r_{a}%
r_{b}r_{i}/r^{4}$, produced by the LIV backgrounds but these Lorentz-violating
corrections do not modify the global asymptotic behavior of the electric field
in (1+2) dimensions: it remains decaying as $1/r$.

From the potential vector (\ref{A02}) we compute the associated magnetic field
produced by a pointlike charge,
\begin{equation}
B\left(  \mathbf{r}\right)  =\frac{q}{2\pi}\epsilon_{ij}\frac{\kappa_{0i}%
r_{j}}{r^{2}}.
\end{equation}
Here, we observe that the LIV parameter $\kappa_{0i}$ engenders an anisotropic
magnetic field whose asymptotic behavior goes as $r^{-1}$. It can be used to
impose an upper-bound for the $\kappa_{0i}$ coefficients by using the
experimental data concerning the two-dimensional physics.

\bigskip For a pointlike charge with velocity $\mathbf{u}$\textbf{,}
$J^{i}(\mathbf{r}^{\prime})=q\delta(\mathbf{r}^{\prime})u^{i}~,\left[
\rho\left(  \mathbf{r}^{\prime}\right)  =0=J\left(  \mathbf{r}^{\prime
}\right)  \right]  $, the scalar potential is%
\begin{equation}
A_{0}\left(  \mathbf{r}\right)  =-\frac{q}{2\pi}\kappa_{0a}u_{a}\ln r~,
\end{equation}
while the vector potential is
\begin{equation}
A_{j}\left(  \mathbf{r}\right)  =-\frac{q}{2\pi}\left[  \left(  1+\frac{1}%
{2}\kappa_{aa}\right)  u_{j}+\frac{1}{2}\kappa_{ja}u_{a}\right]  \ln
r-\frac{q}{4\pi}\kappa_{ab}u_{j}\frac{r_{a}r_{b}}{r^{2}}+\frac{q}{4\pi}%
\kappa_{ab}u_{b}\frac{r_{a}r_{j}}{r^{2}}.
\end{equation}
The respective electric and magnetic field are%
\begin{equation}
E_{i}\left(  \mathbf{r}\right)  =-\frac{q}{2\pi}\kappa_{0a}u_{a}\frac{r_{i}%
}{r^{2}},
\end{equation}%
\begin{equation}
B\left(  \mathbf{r}\right)  =\frac{q}{2\pi}\left[  \left(  1+\frac{1}{2}%
\kappa_{aa}\right)  \epsilon_{ij}\frac{r_{i}u_{j}}{r^{2}}-\epsilon_{ij}%
\kappa_{ab}\frac{r_{a}r_{b}r_{i}u_{j}}{r^{4}}+\epsilon_{ij}\kappa_{ja}%
\frac{3r_{i}u_{a}-r_{a}u_{i}}{r^{2}}\right]  .
\end{equation}

In this model a pointlike scalar source, $J(\mathbf{r}^{\prime})=q_{s}%
\delta(\mathbf{r}^{\prime}),~\left[  \rho\left(  \mathbf{r}^{\prime}\right)
=0=J_{i}\left(  \mathbf{r}^{\prime}\right)  \right]  $, also generates
electromagnetic phenomena whose scalar and vector potentials are given by%
\begin{equation}
A_{0}\left(  \mathbf{r}\right)  =-\frac{q_{s}}{2\pi}S_{0}\ln r,\text{
\ \ }A_{j}\left(  \mathbf{r}\right)  =-\frac{q_{s}}{4\pi}S_{j}\ln
r\text{\ }+\frac{q_{s}}{4\pi}S_{a}\frac{r_{a}r_{j}}{r^{2}},
\end{equation}
leading to the following electric and magnetic field solutions:
\begin{equation}
E_{i}\left(  \mathbf{r}\right)  =-\frac{q_{s}}{2\pi}S_{0}\frac{r_{i}}{r^{2}%
}~,\text{ }B\left(  \mathbf{r}\right)  =\frac{q_{s}}{2\pi}\epsilon_{ij}%
\frac{S_{j}r_{i}}{r^{2}}.
\end{equation}

\subsection{Static solutions for the pure scalar field}

From (\ref{EQSC}) , the stationary solution for the scalar field in can be
expressed as
\begin{equation}
\phi\left(  \mathbf{r}\right)  =\int d\mathbf{r}^{\prime}G\left(
\mathbf{r-r}^{\prime}\right)  J\left(  \mathbf{r}^{\prime}\right)  -\frac
{1}{2\pi}S_{\mu}\int d\mathbf{r}^{\prime}~J^{\mu}(\mathbf{r}^{\prime}%
)~\ln\left\vert \mathbf{r-r}^{\prime}\right\vert \label{A5x}%
\end{equation}
where $G\left(  \mathbf{r-r}^{\prime}\right)  $ is the stationary scalar
Green's function obtained from Eq. (\ref{gsf}), we attain%
\begin{equation}
G(\mathbf{R})=\frac{1}{2\pi}\left(  1+\eta+\frac{1}{2}\kappa_{aa}\right)  \ln
R+\frac{1}{4\pi}\kappa_{ab}\frac{R_{a}R_{b}}{R^{2}}.
\end{equation}
The scalar field generated by a pointlike scalar source, $J(\mathbf{r}%
^{\prime})=q_{_{s}}\delta(\mathbf{r}^{\prime}),$ is
\begin{equation}
\phi\left(  \mathbf{r}\right)  =\frac{q_{_{s}}}{2\pi}\left[  \left(
1+\eta+\frac{1}{2}\kappa_{aa}\right)  \ln r+\frac{1}{2}\kappa_{ij}\frac
{r_{i}r_{j}}{r^{2}}\right]  .
\end{equation}
We thus confirm that scalar field presents a very similar behavior to the one
of the scalar potential, given by Eq. (\ref{A01}).

Similarly, the scalar field produced by a pointlike charge scalar source,
$\rho(\mathbf{r}^{\prime})=q\delta(\mathbf{r}^{\prime})$, and a pointlike
charge with constant velocity $\mathbf{u}$\textbf{,} $J^{i}(\mathbf{r}%
^{\prime})=q\delta(\mathbf{r}^{\prime})u^{i}$,$~$are%
\begin{equation}
\phi\left(  \mathbf{r}\right)  =-\frac{q}{2\pi}S_{0}\ln r~,~\ \phi\left(
\mathbf{r}\right)  =\frac{q}{2\pi}S_{i}u_{i}\ln r,
\end{equation}
respectively, showing similar radial behavior.

\section{Conclusions}

In this work, we have performed the dimensional reduction of the
nonbirefringent CPT-even electrodynamics of the standard model extension. Such
procedure generates a planar Lorentz-violating electrodynamics composed of a
gauge field and a scalar field linearly coupled by a LIV 3-vector $S^{\mu}$.
Both fields have kinetic terms modified by the Lorentz violating symmetric
tensor, $\kappa^{\nu\rho}$. This planar model possesses nine independent LV
components including six parity-even and three parity-odd, being more simpler
than the one of Ref. \cite{DreducCPT}, in which the Lorentz-violation is
governed by 19 parameters (see Lagrangian (\ref{L1})).

The evaluation of the energy-momentum tensor has shown that the density of
energy of the full theory can be positive definite whenever the LV parameters
are sufficiently small. This indicates that the full theory is endowed with
energy stability. The same conclusion is valid for both the pure gauge and the
pure scalar sectors. A complete study on the dispersion relations was
performed. Initially, we have evaluated the dispersion relations of the gauge
and scalar sector (regarded as uncoupled) from the vacuum-vacuum amplitude,
revealing that, at first order, these two fields are described by the same
dispersion relations.\textbf{\ }After, we have carried out the full dispersion
relations, which were exactly computed for some special combinations of the
LIV parameters. The coupling vector $S^{\mu}$ contributes only at second order
for the dispersion relations. All the expressions confirm that the planar
model is nonbirefringent at any order, whereas the original (1+3)-dimensional
model is nonbirefringent only at leading order. From these relations we\ also
conclude the gauge and scalar sector are stable, but endowed with causality
illness. A more careful analysis\ about the physical consistency of this model
(stability, causality, unitarity) is under progress.

We have established the wave equations for the gauge and scalar field and we
have achieved their stationary solutions, via the Green's function technique,
at first-order in the LIV coefficients. The Lorentz-violating terms induce an
anisotropic character to these stationary solutions which now exhibit an
explicit angular dependence. However, the LIV coefficients do not modify the
long distance profile of the solutions, keeping the $r^{-1}$ asymptotic
behavior of the pure Maxwell planar electrodynamics (a fact compatible with
dimensionless nature of the LIV coefficients). The scalar and vector potential
generated by a pointlike scalar charge were carried out as well, showing that
it generates electromagnetic fields. An analogous calculation was accomplished
for the scalar sector, demonstrating that it obeys stationary solutions
similar to the ones of the scalar potential $A_{0}$.

This kind of theoretical framework can find applications in usual planar
systems, such as vortex and Hall systems. At moment, we are particulary
interested in analyzing effects of LIV coefficients in stable vortex
configurations, having already verified that the gauge sector represented by
Lagrangian $\mathcal{L}_{EM},$ when properly coupled to the Higgs sector
endowed with a fourth-order self-interacting potential, supports BPS
(Bogomol'nyi, Prasad, Sommerfeld) solutions. Advances will be reported elsewhere.

\begin{acknowledgments}
The authors are grateful to FAPEMA, CAPES and CNPq (Brazilian research
agencies) for invaluable financial support. The authors also acknowledge the
IFT (Instituto de F\'{\i}sica Te\'{o}rica) staff for the kind hospitality
during the realization of this work.
\end{acknowledgments}

\appendix

\section{The $a_{k}^{(n)}$ coefficients}%

\begin{align}
a_{4}^{\left(  1\right)  }  & =2\kappa_{00},~\ \ \\[0.3cm]
a_{4}^{\left(  2\right)  }  & =\left(  \kappa_{00}\right)  ^{2}+k_{00}%
\text{tr}\left(  \kappa_{ij}\right)  -\mathbf{S}^{2}-\left[  \text{tr}\left(
\kappa_{ij}\right)  \right]  ^{2}+\det\left(  \kappa_{ij}\right)
,~\ \ \ \mathbb{K}=\left[  \kappa_{ij}\right]  ~,\ \\[0.3cm]
a_{4}^{\left(  3\right)  }  & =\mathbb{-\kappa}_{00}\mathbf{S}^{2}%
-\mathbb{\kappa}_{00}\left(  \text{tr}\kappa_{ij}\right)  ^{2}-\kappa
_{ij}S_{i}S_{j}+\left[  \left(  \kappa_{00}\right)  ^{2}+\left(
\det\mathbb{K}\right)  +\mathbf{S}^{2}\right]  \text{tr}\left(  \kappa
_{ij}\right)  .\\[0.3cm]
a_{3}^{\left(  1\right)  }  & =-4\left(  \kappa_{0i}p_{i}\right)
,~\ \\[0.3cm]
a_{3}^{\left(  2\right)  }  & =2S_{0}\left(  \mathbf{S}\cdot\mathbf{p}\right)
+2\left(  \kappa_{0i}\kappa_{ij}p_{j}\right)  -6\kappa_{00}\left(  \kappa
_{0i}p_{i}\right)  ,~\ \\[0.3cm]
a_{3}^{\left(  3\right)  }  & =-2\left(  \kappa_{00}\right)  ^{2}\left(
\kappa_{0i}p_{i}\right)  +2\left(  \mathbf{S}\cdot\mathbf{p}\right)  \left[
k_{00}S_{0}-S_{0}\left(  \text{tr}\mathbb{K}\right)  +\left(  S_{j}\kappa
_{0j}\right)  \right]  +2S_{0}\left(  \kappa_{ij}S_{i}p_{j}\right)  +2\left(
\mathbb{\kappa}_{ia}\kappa_{0a}\right)  \left(  \mathbb{\kappa}_{ib}%
p_{b}\right) \\[0.3cm]
a_{2}^{\left(  1\right)  }  & =2\left(  \kappa_{ij}p_{i}p_{j}\right)
-2k_{00}\mathbf{p}^{2},\\[0.3cm]
a_{2}^{\left(  2\right)  }  & =\left(  \kappa_{00}\right)  ^{2}\mathbf{p}%
^{2}+2\kappa_{00}\left(  \epsilon_{ia}\kappa_{ab}\epsilon_{bj}p_{i}%
p_{j}\right)  -\left(  S_{0}\right)  ^{2}\mathbf{p}^{2}+\left(  S_{i}\right)
^{2}\mathbf{p}^{2}+\left(  \epsilon_{ij}S_{i}p_{j}\right)  ^{2}%
\nonumber\\[-0.3cm]
& \\
& +4\left(  \kappa_{0i}p_{i}\right)  ^{2}-\left(  \epsilon_{ij}\kappa
_{0i}p_{j}\right)  ^{2}+\left(  \mathbb{K}^{2}\right)  _{ij}p_{i}%
p_{j}-2\left(  \epsilon_{ia}\kappa_{ab}\epsilon_{bj}p_{i}p_{j}\right)  \left(
\text{tr}\mathbb{K}\right)  ,\nonumber\\[0.3cm]
a_{2}^{\left(  3\right)  }  & =\left(  \kappa_{00}\right)  ^{3}\mathbf{p}%
^{2}-\left(  \kappa_{00}\right)  ^{2}\left[  2\mathbf{p}^{2}\left(
\text{tr}\mathbb{K}\right)  -\kappa_{ij}p_{i}p_{j}\right]  +\kappa_{00}\left[
2\left(  \epsilon_{ij}S_{i}p_{j}\right)  ^{2}-\kappa_{ia}\kappa_{aj}p_{i}%
p_{j}+4\left(  \kappa_{0i}p_{i}\right)  ^{2}+2\left(  \text{tr}\mathbb{K}%
\right)  ^{2}\mathbf{p}^{2}\right] \nonumber\\[0.3cm]
& -\kappa_{00}\left(  S_{0}\right)  ^{2}\mathbf{p}^{2}-\left(  S_{0}\right)
^{2}\left(  \epsilon_{ia}\kappa_{ab}\epsilon_{bj}p_{i}p_{j}\right)
-2S_{0}\left[  \left(  \kappa_{0i}p_{i}\right)  \left(  S_{j}p_{j}\right)
+\mathbf{p}^{2}\left(  \kappa_{0i}S_{i}\right)  \right]  +\mathbf{p}%
^{2}\left(  \epsilon_{ia}\kappa_{ab}\epsilon_{bj}S_{i}S_{j}\right) \\[0.3cm]
& -\left(  S_{k}\right)  ^{2}\left(  \kappa_{ij}p_{i}p_{j}\right)  -2\left(
\kappa_{0a}p_{a}\right)  \left(  \kappa_{0i}\kappa_{ij}p_{j}\right)  -\left(
\kappa_{0k}\right)  ^{2}\left(  \kappa_{ij}p_{i}p_{j}\right)  -\mathbf{p}%
^{2}\left(  \kappa_{0i}\kappa_{ij}\kappa_{0j}\right)  -\left(  \mathbb{K}%
^{3}\right)  _{ij}p_{i}p_{j}\nonumber\\[0.3cm]
& -2\mathbf{p}^{2}\left(  \text{tr}\mathbb{K}\right)  \det\left(
\mathbb{K}\right)  .\nonumber\\[0.3cm]
a_{1}^{\left(  1\right)  }  & =4\mathbf{p}^{2}\left(  \kappa_{0i}p_{i}\right)
\\[0.3cm]
a_{1}^{\left(  2\right)  }  & =2\mathbf{p}^{2}\left[  \kappa_{00}\left(
\kappa_{0i}p_{i}\right)  -\left(  \kappa_{0i}\kappa_{ij}p_{j}\right)
-S_{0}\left(  S_{i}p_{i}\right)  \right]  -4\left(  \kappa_{0a}p_{a}\right)
\left(  \kappa_{ij}p_{i}p_{j}\right)  ,
\end{align}%
\begin{align}
a_{1}^{\left(  3\right)  }  & =-2\mathbf{p}^{2}\left(  \kappa_{00}\right)
^{2}\left(  \kappa_{0i}p_{i}\right)  -2\kappa_{00}\left(  \epsilon_{ia}%
\kappa_{aj}p_{i}p_{j}\right)  \left(  \epsilon_{bc}\kappa_{0b}p_{c}\right)
+2\mathbf{p}^{2}\left(  S_{0}\right)  ^{2}\left(  \kappa_{0i}p_{i}\right)
+2S_{0}\left(  S_{i}p_{i}\right)  \left(  \kappa_{ij}p_{i}p_{j}\right)
\nonumber\\[0.3cm]
& \ -2\mathbf{p}^{2}S_{0}\left(  \epsilon_{ia}\kappa_{ab}\epsilon_{bj}%
S_{i}p_{j}\right)  +2\left(  \kappa_{ij}p_{i}p_{j}\right)  \left(  \kappa
_{ij}\kappa_{0i}p_{j}\right)  -2\mathbf{p}^{2}\left(  S_{i}\kappa_{0i}\right)
\left(  S_{i}p_{i}\right)  +2\left(  S_{i}p_{i}\right)  ^{2}\left(
\kappa_{0j}p_{j}\right) \\[0.3cm]
& +2\left(  \text{tr}\mathbb{K}\right)  \left(  \epsilon_{ia}\kappa_{aj}%
p_{i}p_{j}\right)  \left(  \epsilon_{bc}\kappa_{0b}p_{c}\right)  +2\left(
\kappa_{0i}p_{i}\right)  \left(  \epsilon_{bc}\kappa_{0b}p_{c}\right)
^{2}.\nonumber\\[0.3cm]
a_{0}^{\left(  1\right)  }  & =-2\mathbf{p}^{2}\left(  \kappa_{ij}p_{i}%
p_{j}\right)  \ ,~\ \ \nonumber\\[0.3cm]
\ a_{0}^{\left(  2\right)  }  & =-\left(  \kappa_{00}\right)  ^{2}%
\mathbf{p}^{4}+\mathbf{p}^{4}\kappa_{00}\left(  \text{tr}\mathbb{K}\right)
+\left(  S_{0}\right)  ^{2}\mathbf{p}^{4}-\mathbf{p}^{4}\left(  \text{tr}%
\mathbb{K}\right)  ^{2}-\mathbf{p}^{2}\left(  \epsilon_{ij}S_{i}p_{j}\right)
^{2}\nonumber\\[-0.2cm]
& \\
& +\mathbf{p}^{2}\left(  \epsilon_{ij}\kappa_{0i}p_{j}\right)  ^{2}+\left(
\kappa_{ij}p_{i}p_{j}\right)  ^{2}+\mathbf{p}^{2}\left(  \text{tr}%
\mathbb{K}\right)  \left(  \kappa_{ij}p_{i}p_{j}\right)  ,\nonumber\\[0.3cm]
a_{0}^{\left(  3\right)  }  & =\mathbf{p}^{4}\left(  \kappa_{00}\right)
^{2}\left(  \text{tr}\mathbb{K}\right)  -\mathbf{p}^{2}\kappa_{00}\left[
\mathbf{p}^{2}\left(  \text{tr}\mathbb{K}\right)  ^{2}+\left(  \epsilon
_{ij}S_{i}p_{j}\right)  ^{2}+\left(  \epsilon_{ij}\kappa_{0i}p_{j}\right)
^{2}\right]  -\mathbf{p}^{4}\left(  S_{0}\right)  ^{2}\left(  \text{tr}%
\mathbb{K}\right)  +\left(  \kappa_{ij}p_{i}p_{j}\right)  \left(
\epsilon_{ij}S_{i}p_{j}\right)  ^{2}\nonumber\\[-0.2cm]
& \\
& +2S_{0}\left(  \epsilon_{ij}S_{i}p_{j}\right)  \left(  \epsilon_{ij}%
\kappa_{0i}p_{j}\right)  -\left(  \kappa_{ij}p_{i}p_{j}-\mathbf{p}%
^{2}\text{tr}\mathbb{K}\right)  \left(  \epsilon_{ij}\kappa_{0i}p_{j}\right)
^{2}-\left(  \kappa_{ij}p_{i}p_{j}\right)  \left(  \kappa_{ij}p_{i}%
p_{j}-\mathbf{p}^{2}\text{tr}\mathbb{K}\right)  \left(  \text{tr}%
\mathbb{K}\right)  .\nonumber
\end{align}

\end{document}